# Broadening the perspective for sustainable AI: Sustainability criteria and indicators for Artificial Intelligence systems

*Friederike Rohde* [a,*1], *Josephin Wagner* [a], *Andreas Meyer* [b], *Philipp Reinhard* [a,c], *Marcus Voss* [b,d], *Ulrich Petschow* [a], *Anne Mollen* [e,f,]

[a] *Institute for Ecological Economy Research, Potsdamer Str. 105, 10785 Berlin, Germany*
[b] *Technische Universität Berlin, Distributed Artificial Intelligence Laboratory, Ernst-Reuter-Platz 7, 10587 Berlin*
[c] *University of Kassel, Research Center for Information System Design, Pfannkuchstraße 1, 34121 Kassel, Germany*
[d] *Birds on Mars GmbH, Marienstraße 10, 10117 Berlin*
[e] *University of Münster, Department of Communications, Bispinghof 9-14, 48143 Münster*
[f] *AlgorithmWatch gGmbH, Linienstr. 13, 10178, Berlin*

***Abstract*** - The increased use of AI systems is associated with multi-faceted social, environmental, and economic consequences. These include non-transparent decision-making processes, discrimination, increasing inequalities, rising energy consumption and greenhouse gas emissions in AI model development and application, and an increasing concentration of economic power. By considering the multi-dimensionality of sustainability, this paper takes steps towards substantiating the call for an overarching perspective on "sustainable AI". It presents the SCAIS Framework (Sustainability Criteria and Indicators for Artificial Intelligence Systems) an assessment framework which contains a set of 19 sustainability criteria for sustainable AI and 67 indicators that are based on the results of a critical literature review, and expert workshops. Its interdisciplinary approach contributes a unique holistic perspective to facilitate and structure the discourse on sustainable AI. Further, it provides a concrete assessment framework that lays the foundation for developing standards and tools to support the conscious development and application of AI systems.

---

[1] * Corresponding author at: Institute for Ecological Economy Research, Potsdamer Str. 105, 10785 Berlin, Germany, E-mail address: friederike.rohde@ioew.de

*Highlights:*

- *Reviews current literature on the social, economic, and environmental impacts of AI*
- *Substantiates the call for an overarching perspective on sustainable AI*
- *Develops sustainability criteria and indicators for AI systems*
- *Introduces the SCAIS framework as an integrated assessment approach for sustainable AI*

## 1. Introduction: Sustainability and Artificial Intelligence

Artificial intelligence systems (AI systems) are complex socio-technical-ecological systems that are associated with multiple social, environmental, and economic challenges. Current discussions raise the question of whether AI systems impede or support a social and ecologically just society [1]. They focus on how AI systems can contribute to various sustainability objectives [2]. Terms such as AI for Earth or AI for social Good (AI4SG) [3] are being used to highlight the potential of AI systems for ecological goals, such as ecosystem monitoring, climate protection, or the energy transition [4–6], sustainable manufacturing [7] as well as for the social good, such as public health issues or education. A growing body of work also covers AI's role in reaching the Sustainable Development Goals (SDGs) [8–11], with an ongoing discussion on whether AI-systems support or impede the SDGs, how those systems should be governed [12] and how reliable and appropriate such assessments are [13]. The discussion regarding the challenges of AI systems in the context of the three sustainability domains [11], is exceeded in a systematic manner in our study, since we aim to broaden the perspective on sustainable AI along the entire life cycle.

The links between AI and the SDGs remain ambivalent because the same technology can be used for conflicting objectives. For example, AI systems can be used through remote-sensing algorithms not only to analyze satellite imagery, gather information on agricultural productivity, or predict the energy consumption of buildings but also to accelerate oil and gas exploration [14].

In contrast to this perspective of AI for Sustainability [2,15], we want to strengthen recent developments in speaking about sustainable AI, or the Sustainability of AI [16–18**], which has already found its way into policy documents [19,20]. We thus want to strengthen perspectives that are more strongly related to the broader social, economic, and ecological impacts, which unfold within a wider social, technological, and environmental context [21]. This sustainable AI perspective addresses the sustainability of developing and using AI systems and brings into focus the need to foster change along the entire lifecycle of AI products [17,18**]. This perspective's necessity is becoming stronger, because AI is not only increasingly important for infrastructures and thus highly relevant to society (e. g. transport, energy, security), but also because "AI itself is becoming an infrastructure that many services of today and tomorrow will depend upon" [3].

Initial work on sustainable AI has focused on singular dimensions of sustainability, with rare exceptions considering two dimensions. Authors have addressed, for example, the environmental costs stemming from the energy needed to train Deep Learning models for natural language processing [18**] or the risk of carbon lock-in that arises with AI becoming an infrastructure [3]. On the social dimension, Halsband et al. [16] examined how intergenerational justice was affected by AI-based decision support for issues with long-term impacts, and current research [22] has developed a guideline that operationalizes and applies social sustainability to the AI development of service robots. On the two-dimensional side, scholars discussed policy implications under the consideration of both environmental (e. g. power consumption of AI systems) and economic aspects (e. g. market concentration in AI markets) of AI development [17]. Recent work from ML researchers [23**] provides a framework similar to this paper, which includes seven categories covering all three dimensions of sustainability and proposing some concrete measures to evaluate AI systems, but with a focus merely on generative AI.

We introduce a comprehensive and novel assessment approach to sustainable AI that takes the multi-dimensionality of sustainability into account (social, ecological and economic, organizational governance dimension) and provides a starting point to govern systemic sustainability risks created by AI [21]. Therefore, this paper develops an overarching perspective on sustainable AI that considers the social, ecological and economic dimensions of sustainability. More specifically, it aims at answering two questions:

1) How can we design an *overarching assessment framework for sustainable AI that takes the AI lifecycle into account?*

2) *Which indicators can be applied to measure the social, ecological, and economic impacts inherent in developing and deploying socio-technical-ecological AI systems?*

We undertake a critical review of literature on social, ecological, and economic impacts related to developing and employing AI systems as well as a review of current assessment approaches for ethical and/or responsible AI. From those reviews, we outline the SCAIS Framework (RQ1) and suggest indicators that reflect discussions of what measures can be taken to fulfill related criteria (RQ2). Our integrative sustainability assessment framework and its set of 19 criteria and 67 indicators allow the sustainability impacts of AI systems to be assessed at concrete, specific levels. These criteria and indicators can be directly implemented within software development standards, processes, and best practices or assessment tools. A comprehensive self-assessment tool for practitioners was developed in 2023[2] within the sustAIn project, which is based on the SCAIS framework and entails those indicators which are assessable through organizations involved in AI development or deployment. However, as the impact levels of AI-systems in figure 1 show, there are some impacts which need to be addressed on a different level, such as policy measures to ensure regulations for recycling of hardware products. It is important to take into account that certain impacts have to be addressed by appropriate policy measures.

The full set of criteria and indicators shown in table 1 aims to capture the whole range of sustainability impacts along the AI lifecycle regarding the social, ecological, economic and governance dimension of sustainability. With this integrative approach, we want to strengthen the perspective of responsible AI development [24,25], which is urgently needed in view of the increasing integration of AI in infrastructures [3] and services of public interest. By putting forward this comprehensive assessment framework for sustainable AI, we aim not only to raise awareness amongst developers, companies, policy makers, and the public but also to provide an assessment framework that enables actors to develop concrete measures to improve AI development and deployment.

---

[2] This tool is available in German language only under: https://sustain.algorithmwatch.org/bewertungstool/ and will subsequently be translated into english.

The remainder of the paper is organized as follows. Section 2 elaborates on our socio-technical approach towards AI and briefly describes the sustainability concept from which our evaluation framework for sustainable AI is derived. Section 3 describes the methodology on how we arrived at the sustainability criteria and indicators for AI systems in detail. The SCAIS framework is introduced in Section 4. Section 5 indicates the challenges for sustainable AI we have identified as well as recommendations for future research.

## 2. An embedded perspective on sustainable AI and the impacts of socio-technical-ecological AI systems

In the following, we explain in more detail our conceptual approach on sustainable AI, which we refer to as the *embedded perspective.* We define AI systems as *systems in which the rules are not defined by humans in the programming of the algorithm but are created by a subsequent learning process (from data). AI systems include both the underlying machine-learning models with inscribed values and the data used for learning.* This study focuses on supervised machine learning as a subset of the AI field while acknowledging other approaches to AI, such as planning and search, knowledge representation, and reasoning. Our interest in that subset stems from the current popularity and increasingly widespread implementation of such machine-learning systems, and the unique challenges arising from the less structured and controlled learning from vast amounts of data. We do not conceive of AI as quasi-objective technology but as embedded in the ideas, interests, knowledge, institutional arrangements of current societies [26] as well as interrelated with the natural environment [17,27] .

Our approach is based on several perspectives. *First,* from a life-cycle perspective on AI systems, we distinguish the following life-cycle phases: 1) Organizational embeddedness; 2) conceptualization; 3) data management; 4) model development; 5) model implementation; 6) model use & decision making [18**].

*Second*, we understand sustainability as a multi-dimensional integrative concept [19] and a normative goal that combines questions of global ecological, social distributional, and inter- and intragenerational justice [20]. Despite its limitations [19] we decided to refer to the three central dimensions, social, ecological, and economic sustainability, to delimit and structure the sustainability criteria. However, this distinction can only be ideal-typical, and a variety of interactions between these dimensions can be observed and are controversially discussed [1**]. Since a just distribution of resources, opportunities and harms is intrinsically linked to economic issues, related questions arising in the context of a socio-ecological transformation [28] often become particularly pressing in the economic context or within economic structures. Consequently, we understand the development and application of AI systems as part of economic activities that must both contribute to maintaining a social foundation by justly distributing opportunities and harms of AI development and usage and ensure compliance with planetary boundaries [29].

Third, we relate this sustainability perspective to the design not only of socio-technical systems [30] but also socio-technical-ecological systems [31]. We conceive of AI systems as social-technical-ecological systems [31,32] because the outcome of those systems can be attributed neither to technology alone, nor to the people who create those systems, nor to the data itself. The socio-technical-ecological characteristics of AI systems derive from the complex interrelations of technical elements and hybrid elements, such as the data model or design process (Figure 1) alongside with interrelations with the natural environment through (direct and embodied) energy and resource use for computing and hardware. Additionally, AI systems are "autonomous, adaptive, and interactive, which means that they acquire many of their features during operation and due to the way they evolve rather than through their initial design" [33] [p. 6]. Rather than focusing on social-ecological (SES) or sociotechnical systems (STS) separately the conceptual framework of social-ecological-technological systems (STES) [31,34] combines both. Initially this conceptual framework was developed as "a heuristic model that provides a critical starting point for positioning and connecting different understandings of different aspects of complex systems" [35] [p. 4] . Our assessment approach follows an understanding of society, technology, and environment as co-constituted and co-emergent entities [31]. To assess and govern systemic risks of AI systems, [21] call for characterizing AI systems as "complex human-machine-ecological systems" [p. 7], implicitly suggesting the integration of social-ecological and social-technical system thinking. Picking up this notion in the context of environmental justice-oriented algorithmic audits, [27] reframe "algorithmic systems as intimately connected to and part of social and ecological systems" [p. 1]. We build on their reflections on ecological-technological couplings of AI systems such as potential impact of the material flow of AI systems along the life cycle or the embodied and shared indirect resource consumption of hardware and infrastructure. The socio-technical-ecological systems perspective on AI calls for the need to recognize that AI systems mediates human–environment relationships in a variety of ways. As we bring together socio-technical and ecological-technological considerations, we distinguish different impact levels of AI-systems, with the different elements and actors, which are part of or able to exert influence on shaping and governing AI systems at the respective levels. This is illustrated in Figure 1. Note that due to the complexity and the multi-dimensionality of the sustainability impacts, this visualization cannot include all aspects relevant for our perspective, such as the AI life cycle. Rather it gives an overview about the impact levels and the entanglements between social, technical and ecological entities that should be addressed when conceptualizing sustainable AI-systems.

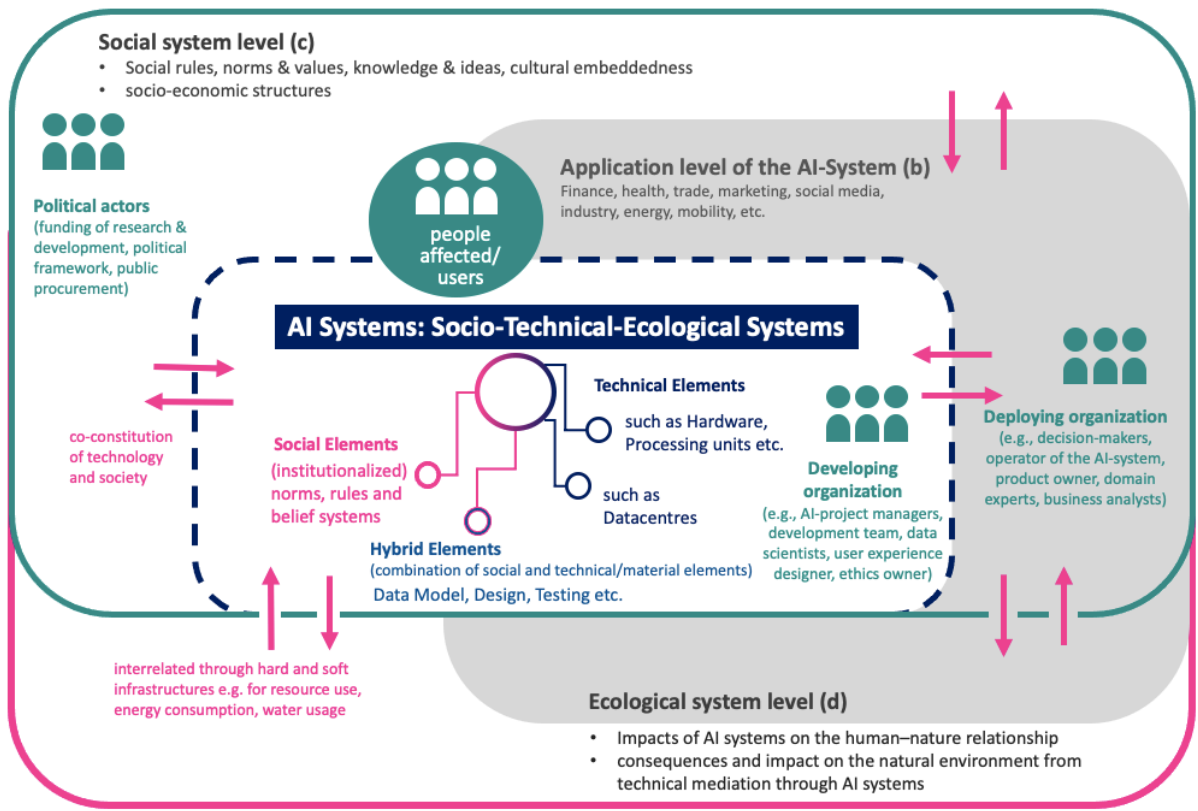

*Figure 1 – AI-systems as Socio-technical-ecological systems, with impact levels and relevant actors for sustainable AI*

The four impact levels that are important for the sustainability assessment of AI systems are:

(a) the level of the social-technical-ecological AI systems as such,

(b) the application level,

(c) the macrosocial level

(d) the ecological system level

The level of the AI system (*a*) primarily addresses AI systems as such, namely the model development, which includes data acquisition and data management, conceptualization and training, testing and possibly retraining, as well as inference. Within the AI lifecycle, different social, technical, and hybrid elements interact and create the AI-system. This system is embedded and interacts with the (macro)-social system (c) and the ecological system (d) and therefore can be understood as a socio-technical-ecological system [27,31,34]. Although we have a strong focus on the AI life cycle, the application level (b) cannot be excluded from a sustainability assessment. Therefore, some of our criteria, such as ecological sustainability potential in the application or effects on the labor market, also relate to the application context, because the effects can only be assessed when considering the respective application. As socio-technical-ecological systems, AI systems are embedded in (macro)social structures (c) and are interrelated to the ecological systems (d) and the natural environment, for example through resource extraction for hardware or through the quantification of nature [36]. We conceive of a co-constitution of technology and society because the development of AI systems is influenced to a large extent by overarching social norms, thought patterns and, of course, the regulatory frameworks – and AI systems simultaneously shape contemporary societies. The social system level includes the structural elements, such as rules and legal frameworks, as well as the cultural embedding and social norms and values. All these aspects influence the knowledge, goals, and framework conditions that shape the development of AI systems.

We apply these conceptual foundations because they can be connected across disciplines and still do justice to the complexity of AI systems and their effects. Our SCAIS framework is not only intended as an impulse to academic discourse but, above all, as a positive contribution to developing AI and to making AI systems more sustainable in the long term in the organizations developing and implementing those systems.

## 3. Methodological approach

Our indicators for sustainable AI are based on a development process which included five steps. The first three steps were part of a conceptual endeavor to arrive at the set of criteria and indicators shown in table 1. The proceeding steps four and five where part of interactions with developing organizations in practice to test the applicability of the indicators and to arrive at an assessment tool.

Conceptual work to establish the set of criteria and indicators in table 1

- Step 1: a critical literature review [37] on social, ecological, or economic impacts of AI systems
- Step 2: Development and deduction of sustainability criteria and criteria profiles
- Step 3: Final operationalization of criteria through measurable and/or assessable indicators

Practical work to test applicability and to make indicators assessable

- Step 4: Grouping of indicators according to their area of application (policy,
- organizational, development) and Pre-Testing & evaluation of indicators relating to organizational execution with AI developing organizations through application and subsequent reflective interviews
- Step 5: Development of an online self-assessment tool

The *first step* of our approach included a critical literature review as well as a review of current assessment approaches for ethical and/or responsible AI. We choose this method because the issue under study is of a complex or heterogeneous nature and relates to various academic discourses such as AI ethics, green AI, life-cycle assessment, or sustainable AI. Accordingly, our review includes a broad and representative but not necessarily comprehensive set of articles [38]. Our review included academic papers published between 2019 and 2021 that contained research about the social, ecological, or economic impacts of AI-based systems and issues concerning the organizational governance. The aim of our review is to critically analyze extant literature on the emerging issue of sustainable AI and to reveal weaknesses, controversies [38] and the relevant sustainability impacts discussed. According to the sustainability dimensions – social (most discussed in AI ethics), ecological and economic – we searched for terms based on the combination of artificial intelligence (including deep learning, machine learning etc.) and the corresponding dimensions. We started with overview papers and most-cited papers and identified further studies through backward and forward search by screening titles and abstracts. We read the abstracts of the identified studies to determine their relevance to the research topic. We applied the following inclusion criteria for our search: the use of machine learning or deep learning, measurability of discussed themes, connection to the definition and dimensions of sustainability and a relevance of the discussed issues with regard to social, economic or ecological impacts. The last criterium is important because not all the impacts, that can or may occur from AI-Systems are discussed with reference to sustainability. However, we argue that an overarching sustainability perspective should address all those impacts, including economic impacts such as market concentration. This overarching perspective allows address a wide range of sustainability impacts and provides a path to identify possible trade-offs, as already indicated in the research [1**].

We skimmed through the full-text articles to evaluate further the quality and eligibility of the studies. The overall sample consisted of 112 academic publications. Beyond that we included eight AI impact assessments in our analysis which are not primary published in academic papers but available through websites or online repositories, such as the Responsible AI Design Assistant[3], the Algorithmic Impact Analysis of the Government of Canada[4] or the AI Ethics label from the AI Ethics Impact Group[5].

In the *second step* we developed an initial set of categories for social, ecological, and economic impacts and problems discussed in the literature. Our interdisciplinary project team clustered and refined the categories incrementally to arrive at our final set of sustainability criteria. The final criteria are aimed at offering an approach which is detailed enough for a comprehensive assessment but not overly complex and thus still remain applicable for AI developing or implementing organizations. After arriving at the criteria, we added the dimension of organizational governance. It became apparent that many of the desirable developments [35] from a sustainability perspective are related to the way AI development is organized, monitored and the framework conditions under which it takes place. Once the criteria were developed, we wrote criteria profiles for each criterion, which contained a description of the criterion and the AI-specific impacts as well as possible or existing indicators that allow for measuring sustainability.

In a *third step* we refined the operationalization of the criteria and developed or deducted the final indicators. We arrived at the operationalization through approaches that were apparent in the existing literature, through existing assessment approaches or we developed new indicators within the project team. Especially for the economic consequences of AI systems, there was little literature providing indicators. Consequently, from impacts and questionable developments described in the literature and comparable economic indicators, the project team developed specific criteria and indicators to assess those impacts. Accordingly, the indicators were either proposed by the literature explicitly, derived from the literature implicitly or taken from the eight AI assessments that we consulted. Consequently, we present a novel set of sustainability criteria and indicators, which does not yet exist in the scholarship on AI and sustainability.

---

[3] https://oproma.github.io/rai-trustindex/
[4] https://open.canada.ca/aia-eia-js/?lang=en
[5] https://www.ai-ethics-impact.org/en

To be able to make the criteria and indicators as useful as possible for practitioners, i.e. organizations developing and implementing AI systems in a *fourth step* we tested and evaluated the criteria and indicators for their applicability in practice. As a preparatory step we re-aggregated the identified criteria and indicators into three categories with regards to their level of applicability: 1) applicable on organizational level (e.g. existence of an organizational AI Code of Conduct) 2) applicable to the development of specific AI systems (e. g. use of quality standards for data sets) 3) applicable to policy, regulatory or governance level (e. g. open data policies beyond individual or organizational control). Indicators from group 1 and 2 were then integrated into a questionnaire and pre-tested with four AI-developing organizations, with an initial insight interview, an application phase of the questionnaire and a post-application feedback interview, which then led to a revised questionnaire. These indicators were further discussed in two expert workshops, namely *sustainable AI labs*, in May 2021 and February 2022 with experts from industry, policy, and academia

In a *fifth step* based on the above consultation we developed a self-assessment tool consisting of a maximum of 66 questions that helps organizations assess their organizational handling of AI systems regarding our sustainability indicators. The tool is based on a point-based evaluative framework, which aggregates participant's responses along 10 groups, allowing for detailed reflection of the results and for formulating precise recommendations for improvement.

## 4. Sustainability criteria and indicators for AI systems – The SCAIS Framework

From the literature reviews, we obtained six cross-sectional criteria and thirteen further sustainability criteria directly related to one of the dimensions. For these criteria, we developed 67 indicators (Table 1). The indicators' column of Table 1 reflects currently discussed concepts regarding what measures and actions can be taken to fulfill a particular criterion. Since a comprehensive discussion of each indicator exceeds the scope of this paper, we describe the indicators in the table to provide an overview of our approach, which is open to refinement and discussion in future research on sustainable AI.

*Table 1: The SCAIS Assessment Framework (Sustainability Criteria and Indicators for Artificial Intelligence Systems).*

| Criteria | Indicators (Operationalization) | Life Cycle Phase | References |
|---|---|---|---|
| **(Organizational) governance dimension** (Cross-cutting criteria) | | | |
| (1) Defined responsibilities | (1) There are contact persons for ethical and social matters<br>(2) The allocation of responsibility is clearly and transparently regulated & documented<br>(3) There are regulations on liability aspects | 1 | [39,40] |
| (2) Code of conduct | (4) Norms and values for the implantation and use of AI systems defined in a code of conduct | 1 | [39,41] |
| (3) Stakeholder participation | (5) Identification and classification of stakeholders<br>(6) Integration of stakeholders into design, test and release processes | 1,2,4,5,6 | [39,42] |
| (4) Documentation | (7) Documentation of information regarding objectives, domain, users, data, model, feature selection, inputs, tests, metrics, etc. (Model card) | 1,2,3,4,5,6 | [39,42,43] |
| (5) Risk management | (8) Implementation of risk assessment<br>(9) Implementation of risk monitoring<br>(10) Implementation of risk management | 1,2,4,6 | [42,44] |
| (6) Complaint mechanism | (11) Option to report errors, unfair and discriminatory decisions, privacy intrusions, etc. to AI-operating company | | |
| **Social dimension** | | | |
| (7) Transparency & accountability | (12) Parameter count<br>(13) AI type (deep learning vs. statistical learning)<br>(14) Use of methods for increasing transparency & explainability<br>(15) Information about AI usage available<br>(16) Access to information about functionality | 1,3,5,6 | [24,39,42,43,45–49*] |
| (8) Non-discrimination & fairness | (17) Assessment of the potential for discrimination<br>(18) Usage of methods for measuring fairness and bias<br>(19) Definition of vulnerable groups and protected attributes<br>(20) Measures to eliminate discrimination | 1,3,4,5,6 | [44,49*–52] |
| (9) Technical reliability & human supervision | (21) Mechanisms for performance control<br>(22) Ensuring appropriate data quality<br>(23) Opportunity for human control | 3,5,6 | [53–58] |
| (10) Self-determination and data protection | (24) Privacy-by-design<br>(25) Users have control over their data<br>(26) Earmarked data use<br>(27) Notifications regarding data use<br>(28) Self-motivated use of AI-systems<br>(29) Abandonment of addiction-enhancing mechanisms (nudging, dark patterns) | 2,3,6 | [2,5,44,49*,59] |
| (11) Inclusive & participatory design | (30) Applying co-design principles<br>(31) Ensuring accessibility | 2 | [44,60,61] |
| (12) Cultural sensitivity | (32) Team diversity | 1,2,5 | [60,62–64] |
| | (33) Integration of local experts and natives | | |
| | (34) Transferability of the AI system to adapt to local and new application contexts, norms, and values | | |
| *Ecological dimension* | | | |
| (13) Energy consumption | (35) Energy consumption is considered during the system development | 1,4,6 | [65–71] |
| | (36) Models with lower complexity are favored during model selection | | |
| | (37) Pre-trained models and transfer learning are used | | |
| | (38) Parameters that capture the model efficiency are measured | | |
| | (39) Methods for model compression are used | | |
| | (40) Methods for efficient training of the models are applied | | |
| | (41) Measures are used to reduce the amount of data | | |
| (14) $CO_2$ and GHG emissions | (42) $CO_2$ footprint | 1,3,4,5,6 | [67,72–79] |
| | (43) $CO_2$ efficiency | | |
| | (44) Emission compensation | | |
| (15) Sustainability potential in application | (45) Sustainable target function | 1,2,3,6 | [5,14,46,72,80] |
| | (46) Consideration of sustainability criteria in decision systems | | |
| | (47) Promotion of sustainable products | | |
| | (48) Promotion of sustainable consumption or sustainable consumption patterns | | |
| | (49) Reduction of resource consumption of processes or products | | |
| | (50) Impact of the AI system on the product quality and service life | | |
| (16) Embodied & shared resource consumption of hardware infrastructure | (51) Certified hardware (energy & resource efficiency) | 1,2 | [72,81,82] |
| | (52) Certified data center (transparency, energy & resource efficiency) | | |
| | (53) Efficiency metrics for data centers (e.g. Power-/Water-/Carbon usage effectiveness) | | |
| | (54) Hardware recycling rate | | |
| | (55) Hardware reuse rate | | |
| | (56) Use of waste disposal scenarios for hardware | | |
| *Economic dimension* | | | |
| (17) Market diversity and exploitation of innovation potential | (57) Accessibility of code | 1,3,4,5,6 | [83–85] |
| | (58) Accessibility of data (data pools) | | |
| | (59) Accessibility of AI tools | | |
| | (60) Interfaces (APIs) | | |
| | (61) Multihoming & Compatibility | | |
| (18) Distribution effect in target markets | (62) Adaptability to data volumes and action requirements | 1,2,5,6 | [83] |
| | (63) No differences in accuracy between major and marginalized market players | | |
| | (64) Diversity of employing customers | | |
| | (65) Support for SMEs & NGOs | | |
| (19) Working conditions and jobs | (66) Evaluation of effects on working conditions | 1,6 | [86*–89] |
| | (67) Fair wages along the AI-lifecycle | | |

*Legend of the AI life-cycle phases: 1) Organizational embeddedness; 2) conceptualization; 3) data management; 4) model development; 5) model implementation; 6) model use & decision making*

Our SCAIS framework addresses the diversity of impact levels defined above while providing practical value and guidance.

In the following we want to illustrate how our self-assessment tool can be used in practice through describing a *hypothetical in-depth-scenario*. As our pre-testing has demonstrated the self-assessment tool developed based on the above criteria helps organizations, who are developing or using AI systems, to learn how well positioned they are regarding the sustainability of their AI systems and where there is room for improvement[6]. Considering a "Code of Conduct" as a sustainability criterion in an organizational governance dimension, the questionnaire would be asking for the existence of such a Code of Conduct and list several "norms and values for the implementation and use of AI systems defined in the code of conduct" (cf. Table 1, Code of Conduct on indicator level). An organization would for instance be made aware that it is a good first step to have established a Code of Ethics that establishes certain principles such as transparency, non-discrimination, human oversight etc. as orientation for any AI development. But it could improve if the adherence to such a Code of Ethics would be monitored by an internal oversight body, which also serves as a central resource for information and has the power to object in AI development processes. It would learn what information about AI models and data sets should be made public and how purposefully choosing less complex models could help with effective oversight. If the organization develops AI systems with direct impact on people, for instance through automated decision-making, the self-assessment tool would for instance evaluate positively when the organization involves marginalized actors as stakeholders in consultation processes on the system's impact across its lifecycle. It would suggest that an organization could improve when the relevant workforce is very homogeneous (e. g. age, gender) and no diversity management is in place. The tool recognizes if an organization usually deploys carbon efficiency methods in AI development and uses eco-certified hardware and data centers, but it could improve when it so far does not cooperate with recycling or re-manufacturing companies when disposing of old hardware. In summary, based on the SCAIS framework the self-assessment helps organizations developing or implementing AI systems to enhance their approaches and technologies step by step throughout the entire ML life cycle[7].

---

[6] Since we have done the pre-testing under the agreement that the obtained data will not be published, we are in the following presenting fictitious examples of what insights the self-assessment tool delivers.

## 5. Challenges and implications for AI development, research, and policy

Previous discourses around sustainable AI focused strongly on the ethical and environmental dimensions, with the ethical discourse mostly limited to presenting principles [90]. With our work, we highlight the need to take a holistic approach to the sustainability of AI and use an indicator-based approach to show possibilities for the practical implementation of sustainable AI systems. In doing so, we go beyond discussing mere principles. However, our work reveals two main challenges that need to be overcome on the way to sustainable AI:

(1) Practical implications: a need for regulation at the policy level and industry standards
(2) Conceptual implications: a need for research and reflection on the interdependencies between sustainability-related impacts.

First, from a practical perspective, there is ***a*** *need for regulation and industry standards.* One of the great obstacles to making AI more sustainable is a lack of data and adequate documentation processes during AI development and deployment. This lack can be rooted in insufficient awareness about the risks and negative impacts among AI-development communities [60] or to organizational peculiarities arising from time constraints in development teams [91]. However, our comprehensive set of sustainability criteria can serve as a source for companies developing AI systems, as well as for those deploying systems, and can strengthen awareness of wider sustainability impacts related to AI systems. Since policy approaches [44,92,93], standards [94] and certification initiatives [95] are increasingly integrating sustainability aspects, the current momentum needs to result in changing practices amongst developers and mandatory reporting requirements. Policy approaches that take sustainability as an overarching principle into account should address all sustainability dimensions and thus shape the increasing impacts of AI systems [21] in favor of people and the planet. This study with its criteria and indicators can serve as a basis for such policies, standards, or industry best practices.

---

[7] Step-by-Step guidelines can be assessed through the project website: https://sustain.algorithmwatch.org/en/step-by-step-towards-sustainable-ai/

Second, we identify conceptual implications, in particular the need for research on and reflection of the entanglements and interrelatedness of the sustainability impacts of AI systems and the systemic risks [21]. Because AI systems are socio-technical-ecological-systems , they possess complexity and embeddedness, and their impact levels are thus diverse and interdependent. To achieve a comprehensive assessment approach, more focus should be placed on those interdependencies. For example, moving more computation from on-premise data centers to the cloud could benefit the ecological dimension due to the often greater energy efficiency and use of cloud data centers' hardware [78], but it also might give additional momentum to the concentration on the cloud market [96]. Thus, improving AI systems regarding one sustainability dimension may, at times, lead to risks for another sustainability dimension. On the other hand, it is also conceivable that synergies can arise. Future research should therefore assess interdependencies between the impacts of AI systems.

The set of criteria presented can serve as a starting point for assessing those impacts and reveal possible conflicting objectives. However, the indicators that we developed do not fully cover all aspects of the Socio-Technical-Ecological-Systems Framework. For example, they lack considerations of how technology interacts with the social-ecological system and not only with the social system or the ecological systems. Our approach covers social-technical and ecological-technical couplings but the whole integration of social, ecological, and technological interrelations with regard to AI systems provides rich potential for further research. Furthermore, critical discussion, learning and organizational sensemaking [8] should be strengthened through a comprehensive sustainability assessment. In addition to the pure description of interdependent impacts of AI systems, however, there must also be a social negotiation process on how to deal with (partly irresolvable) trade-offs. In future debates on the use of AI systems, these trade-offs between different sustainability impacts will play a major role and need to be investigated more thoroughly in future research. Achieving sustainability with regard to the AI-lifecycle is multi-dimensional and will require further research and societal negotiation to define which of the sustainability criteria should be given higher priority and which impacts should be addressed with increased attention. This study can support such research by providing a holistic assessment framework of the pertinent dimensions.

## 6. Conflict of interest statement

Nothing declared.

## 7. Acknowledgment

We thank AlgorithmWatch, for the inspiring work in the *SustAIn* Project and Lina Engel from IOEW for supporting the editing process. This work was financially supported by the German Federal Ministry for the Environment, Nature Conservation, Nuclear Safety and Consumer Protection under grant numbers 67KI2060B, 67KI2060C.

## 8. References and recommended reading

discourses around sustainable AI and brings into focus the need to foster change in the entire life cycle of AI products.

Instead, the authors highlight industry's responsibility to consider the impact of the technologies it is shaping on labor markets and employment opportunities, introducing the notion of 'shared prosperity'. They lay out a framework for systematically evaluating this impact.